\documentclass{article}
\usepackage{amsmath}
\usepackage{amssymb}

\input{tcilatex}
\begin{document}

\title{Multifractality in the general cosmological solution of Einstein's
equations}
\author{John D. Barrow \\
DAMTP, Centre for Mathematical Sciences,\\
University of Cambridge,\\
Wilberforce Road, Cambridge CB3 0WA\\
United Kingdom\ \ \ \ \ \ \ \ }
\maketitle
\date{}

\begin{abstract}
We demonstrate the scale invariance of the vacuum Bianchi type IX equations
and use this to argue for the possibility of multifractal turbulence as a
realisation of the suggestion by Belinski that there will be a fragmentation
of local regions of inhomogeneous Mixmaster chaos on approach to an initial
inhomogeneous cosmological singularity. Differences between the
gravitational and hydrodynamical situations are outlined. Various potential
obstacles to this picture of gravitational turbulence are discussed together
with links to preinflation.
\end{abstract}

\section{Introduction}

We have recently explored the potential effects of synchronisation of
Mixmaster oscillations between different regions of an inhomogeneous
Mixmaster universe on approach to the initial cosmological singularity \cite%
{synchro}. Here, we consider another generalisation of the standard picture
of Bianchi type IX evolution towards the singularity in the inhomogeneous
generalisation. Belinski \cite{bel} has suggested that there could be
further added structure to the general inhomogeneous Mixmaster oscillations
in the form of the transport of energy from large to small scales leading
(like in any complex nonlinear system with infinite number of degrees of
freedom) to the endless generation of excitations of smaller and smaller
spatial scales. This process can be called the fragmentation of the
Mixmaster oscillatory regime and its rise is supported by the mathematical
fact of the monotonic increase of spatial gradients of the metric on
approach to the singularity that was observed by Kirillov and Kochnev \cite%
{kir, kir2} and by Montani \cite{mont, mont2}. A further discussion of this
fragmentation idea appears in the book \cite{bel2}, see also ref. \textbf{%
\cite{JBrep}.} \ 

This phenomenon is reminiscent of three-dimensional fluid turbulence but
with no minimum dissipative scale and so the energy transport continues down
to zero scale. The approach to the singularity at $t=0$ produces an infinite
number of subdivisions of each local inhomogeneous Mixmaster dynamical
region and fragmentation continues ad infinitum. There will only be a
cut-off to this process and to the downward energy transfer if there is a
minimum length scale where dissipation of the gravitational wave energy can
occur; by particle production of gravitons, or other degrees of freedom; or
if there is a bounce \cite{BGang, BGang2} at a minimum non-zero radius; or
if the oscillations cease because of asymptotic dominance of the local
dynamics by a massless scalar field or the kinetic energy of a\ scalar field
with a non-zero potential.

Another ingredient that is potentially relevant, and maybe even related, are
spikes and their oscillations, \cite{lim, lim2}. In the next section we
provide some more rigorous basis for our

\section{Type IX scaling and multifractal behaviour}

The spatially homogeneous diagonal Bianchi IX metric is,

\begin{equation}
ds^{2}=dt^{2}-\gamma _{ab}(t)e_{\mu }^{a}e_{\nu }^{b}dx^{\mu }dx^{\nu },
\label{met}
\end{equation}

where

\begin{equation}
\gamma _{ab}(t)=diag[a^{2}(t),b^{2}(t),c^{2}(t)],  \label{met2}
\end{equation}%
and

\begin{equation}
e_{\mu }^{a}=%
\begin{pmatrix}
\cos z & \sin z\sin x & 0 \\ 
-\sin z & \cos z\sin x & 0 \\ 
0 & \cos x & 1%
\end{pmatrix}%
.  \label{met3}
\end{equation}

The general relativistic field equations in vacuum Bianchi type IX with
scale factors $a(t),b(t),c(t)$ in comoving proper time $t$, \cite{LL, bel2,
mis}, are:

\begin{equation}
\left( \dot{a}bc\right) ^{\cdot }=\frac{1}{2abc}\left[ \left(
b^{2}-c^{2}\right) ^{2}-a^{4}\right] ,  \label{1}
\end{equation}%
\begin{equation}
\left( a\dot{b}c\right) ^{\cdot }=\frac{1}{2abc}\left[ \left(
a^{2}-c^{2}\right) ^{2}-b^{4}\right] ,  \label{2}
\end{equation}%
\begin{equation}
\left( ab\dot{c}\right) ^{\cdot }=\frac{1}{2abc}\left[ \left(
a^{2}-b^{2}\right) ^{2}-c^{4}\right] ,  \label{3}
\end{equation}

\begin{gather}
(\frac{\dot{a}}{a}+\frac{\dot{b}}{b}+\frac{\dot{c}}{c})^{2}=\left( \frac{%
\dot{a}}{a}\right) ^{2}+\left( \frac{\dot{b}}{b}\right) ^{2}+\left( \frac{%
\dot{c}}{c}\right) ^{2}+  \label{4} \\
+\frac{1}{2a^{2}b^{2}c^{2}}%
(a^{4}+b^{4}+c^{4}-2a^{2}b^{2}-2a^{2}c^{2}-2b^{2}c^{2}).  \notag
\end{gather}

\bigskip\ Eqn. (\ref{4}) is also expressed more simply before integration as,

\begin{equation}
\frac{1}{a}\frac{d^{2}a}{dt^{2}}+\frac{1}{b}\frac{d^{2}b}{dt^{2}}+\frac{1}{c}%
\frac{d^{2}c}{dt^{2}}=0.  \label{A}
\end{equation}

Now consider the scaling behaviour of these equations by a power, $h$, of a
constant positive scaling factor $A$. We set

\begin{eqnarray}
(a,b,c) &\rightarrow &A^{h}(a_{\ast },b_{\ast },c_{\ast }),  \label{5} \\
t &\rightarrow &A^{h}t_{\ast }.  \label{6}
\end{eqnarray}

Then, eqn. (\ref{1}) transforms in the $\ast $ variables to (where $^{\prime
}$ denotes $d/dt_{\ast }$),

\QTP{Body Math}
\begin{equation}
A^{h}(a_{\ast }^{\prime }b_{\ast }c_{\ast })^{\prime }=\frac{A^{h}}{2a_{\ast
}b_{\ast }c_{\ast }}\left[ \left( b_{\ast }^{2}-c_{\ast }^{2}\right)
^{2}-a_{\ast }^{4}\right] ,  \label{B}
\end{equation}%
for arbitrary constants, $A$ and $h$. The same invariance holds for eqns.(%
\ref{2})-(\ref{3}) by cyclically permuting letters $a,b$,and $c$. The first
integral,eq.(\ref{4}), transforms as,

\begin{gather}
\frac{1}{A^{2h}}\left( \frac{a_{\ast }^{\prime }}{a_{\ast }}+\frac{b_{\ast
}^{\prime }}{b_{\ast }}+\frac{c_{\ast }^{\prime }}{c_{\ast }}\right) ^{2}=%
\frac{1}{A^{2h}}\left[ \left( \frac{a_{\ast }^{\prime }}{a_{\ast }}\right)
^{2}+\left( \frac{b_{\ast }^{\prime }}{b_{\ast }}\right) ^{2}+\left( \frac{%
c_{\ast }^{\prime }}{c_{\ast }}\right) ^{2}\right]  \label{9} \\
+\frac{1}{2a_{\ast }^{2}b_{\ast }^{2}c_{\ast }^{2}A^{2h}}(a_{\ast
}^{4}+b_{\ast }^{4}+c_{\ast }^{4}-2a_{\ast }^{2}b_{\ast }^{2}-2a_{\ast
}^{2}c_{\ast }^{2}-2b_{\ast }^{2}c_{\ast }^{2}),  \notag
\end{gather}%
and again we see it has the same invariance, as expected, because eq. (\ref%
{4}) is a first integral of eqns.$\ $(\ref{1})-(\ref{3}), and the equation
is scale invariant under all $A$ and $h$ scaling transformations. The same
scaling is seen more obviously from eqn. (\ref{A}). The 'toy' type IX model
created by Fleig and Belinski \cite{bfr} does not have this scaling
property.{} This is not surprising as the equations were created by setting $%
c(t)=0$ and so the scale invariance in eq. (\ref{5}) is obviously broken.

The scale invariance of these equations has important consequences. Parisi
and Frisch \cite{PF,Ben,fr} famously considered a similar problem in the
context of scale invariance of the Euler and Navier-Stokes equations and its
relation to the presence of turbulence with multifractal structure. The
scale invariance property of the vacuum type IX equations we have shown
means that in the inhomogeneous generalisation of the Mixmaster model there
can exist many different values of $h$, each occurring say with some
probability $P_{r}(h)$ on any fixed scale $r$. So, a turbulent situation
appears as a superposition of many different scale-invariant flows. In order
to \ maintain scale invariance when averaging correlation functions one need 
$P_{r}(h)\simeq r^{f(h)}$, for some unknown function $f(h)$ so that the
scale invariance holds for the distribution $P_{r}(h)$ also. Parisi and
Frisch took \ $f(h)$ to be $3-D(h)$, with $D(h)$ taken to be the fractal
(Hausdorff) dimension of the scale invariant solution with scaling exponent $%
h$. Hence, this scenario is referred to as multifractal turbulence and we
see that it arises in the inhomogeneous extension of the\textbf{\ }vacuum
Bianchi type IX equations. In the hydrodynamic problem, a further constraint
is introduced by using the Kolmogorov constraint that there is
scale-invariant energy flow per unit time from large to small scales. This
requires constant energy flow rate, $v^{2}/t\simeq v^{3}/r,$ through scale $%
r $, so $v\varpropto r^{1/3\text{ \ }}$independent of the scaling
parameters: the famous Kolmogorov spectrum.

We know that a turbulent inertial region requires a lower length limit
cut-off where dissipation of the energy injected on large scales can occur.\
Of course, in the hydrodynamical case this scale breaks the scale invariance
but the situation may be simpler in the cosmological case as it is not
obvious if a lower length scale exists as a cut-off to the turbulent
fragmentation. There are three simple possibilities:

a. The lower length cut-off is the Planck scale, $l_{pl}\simeq 10^{-33}cm$.
However, if this is the case there will be a very small number of
oscillations (of order 10) between the present and the Planck time (a mean
expansion scale factor change of order $10^{60}$) as they occur in $ln(\ln
t) $ time -- far too few to set up a well-developed inertial regime for some
type of gravitational turbulence.

b. The fragmentation degenerates into the formation of some organized
objects, for example, as gravitational solitonic structures in which there
is a balance between gravity and dispersion, which act a limiting case \cite%
{belzak, verd}. They may also be related to the synchronisation process
outlined in ref. \cite{synchro}. This scenario also appears in other aspects
of hydrodynamic turbulence studies. The Kolmogorov-Hinze theory of the
fragmentation of droplets in a chaotic turbulent flow is relevant \cite{kol,
hinz},. Using Kolmogorov's classic velocity spectrum, they derived a
criterion for the maximum size of droplet that will not undergo
fragmentation when the turbulent flow intensity exceeds the surface tension 
\cite{perl}. Many theoretical and experimental studies have been made of the
details of this process and the generalisations and limitations of the
Kolmogorov-Hinze picture \cite{east}. It is suggestive of an analogy in the
cosmological picture where Weyl curvature oscillations overcome self-gravity
to perpetuate anisotropic stretching and fragmentation of regions.

b. There is no lower limit and the chaotic oscillations and turbulent
behaviour continue all the way to $t=0$. This results in an infinite number
of spacetime oscillations of the scale factors $a,b,c$ in $t$ time (or
equivalently $a_{\ast },b_{\ast ,}c_{\ast }$ in $t_{\ast }$ time) but with
the conceptual problem of what is the meaning of the model at times earlier
than the Planck time when less than one photon is present inside the horizon
and so statistical mechanics is meaningless (and perhaps the concept of
space and time oscillations as well (but see ref.\cite{misner}) in a
non-quantum gravitational form.

We should also mention the potential links to the study of the
pre-inflationary structure of the very early universe, see for example ref. 
\cite{mat}. Preinflation can imprint particular fluctuation scales when
density perturbations cross the horizon and provide a way to access
information about the preinflationary (possibly chaotic) structure of the
very early universe. Our study has focussed on the multifractal evolution on
approach to an initial singularity ($t\rightarrow 0$) and so any remnant of
preinflation has to look at the evolution in the opposite time-sense where
the fragmentation process will run in reverse. Again, a we have stressed
above, any interval of Mixmaster evolution not including $t=0$ creates a
finite number of oscillations and fragmentations and so as $t$ grows there
may be a residual effect of early fragmentation but it will not have been in
the fully developed gravitational turbulence range. This may repay further
investigation. 

In this paper we have tried to provide some simple underpinning for the idea
that there is random fragmentation of Mixmaster dynamics in different
regions on approach to the singularity in an inhomogeneous version of the
Bianchi type IX universe when the inhomogeneities are still not too large.

\textbf{Acknowledgements}. I would like to thank V.A. Belinski for
encouragement, discussions and detailed comments. Support by the Science and
Technology Facilities Council (STFC) of the UK is acknowledged.

\end{document}